\documentclass{article}

\def\xxinput#1{\input#1}

\xxinput{vsolj02.sty}

\usepackage[dvipdfmx]{graphicx}
\usepackage[comma,colon]{natbib}

\usepackage[OT2,T1]{fontenc}

\def\cite{\citealt}
\setcitestyle{aysep={}}

\newcounter{author}
\def\altaffilmark#1{$^{#1}$}
\def\altaffiltext#1{$^{#1}$\,}

\def\authorcount#1#2{{\refstepcounter{author}\label{#1}
                     \altaffiltext{\ref{#1}}{#2}}}

\begin{document}

\begin{center}

\title{WFI J161953.3$+$031909: eclipsing ER UMa-type and Z Cam-type star}
\vskip -2mm
\title{in the period gap}

\author{
        Taichi~Kato\altaffilmark{\ref{affil:Kyoto}}
}
\email{tkato@kusastro.kyoto-u.ac.jp}

\authorcount{affil:Kyoto}{
     Department of Astronomy, Kyoto University, Sakyo-ku,
     Kyoto 606-8502, Japan}

\end{center}

\begin{abstract}
\xxinput{abst.inc}
\end{abstract}

\section{Introduction}

   WFI J161953.3$+$031909 was discovered by \citet{rau06j1619}
during a wide-field search for orphan afterglows of
gamma-ray bursts (GRBs).  The object showed sudden brightening
by 2.4~mag between 1999 June 14 and June 19.  Although
\citet{rau06j1619} described that the object decayed over 50--90~d
before returning to quiescence, this figure was based on
very sparse observations (see their figure 5).
\citet{rau06j1619} suggested it to be dwarf nova.
Based on detections of 9 X-ray photons with ROSAT, they derived
a high $L_{\rm X}/L_{\rm opt}$=0.6 and was considered to be
consistent with an SU UMa star referring to \citet{ver97ROSAT}.
\citet{rau06j1619} derived a distance of order of
a few hundred parsec based on this X-ray observation.

   \citet{rau07DNe} studied this object in more detail
in 2006 and found it to be an eclipsing binary with
a period of 0.099041(9)~d.  \citet{rau07DNe} also detected
strong Balmer and He~I emission lines.  \citet{rau07DNe}
also attributed a broad feature as a Bowen blend,
although there was no feature of He~II emission, which
should be present when the Bowen blend is present.
\citet{rau07DNe} interpreted that WFI J161953.3$+$031909
has a low mass-transfer rate expected for an object
in the period gap based on the absence of an orbital hump
or the asymmetry in the eclipse.  Based on this interpretation,
\citet{rau07DNe} concluded that the ``outburst'' detected
in 1999 \citep{rau06j1619} represented the mean out-of-eclipse
brightness rather than a dwarf nova-type outburst and
that the faint ($R$=19.9) observations corresponded to
eclipses.  \citet{rau07DNe} concluded that there was
no evidence for a dwarf nova outburst in this system
and corrected $L_{\rm X}/L_{\rm opt}$ to be $\sim$0.1
using the out-of-eclipse brightness rather than what
had been considered to be quiescence.

   Based on the classification by \citet{rau07DNe},
the AAVSO Variable Star Index
\citep{wat06VSX} classified this object to be an eclipsing
novalike object (at the time of this writing on 2022 March 13),
although it was apparently inconsistent with
the low mass-transfer rate as stated by \citet{rau07DNe}.
This object was also detected as a transient Gaia19cwd\footnote{
  $<$http://gsaweb.ast.cam.ac.uk/alerts/alert/Gaia19cwd/$>$.
} by the Gaia Photometric Science Alerts Team.
The light curve on the page of Gaia19cwd showed scattered
magnitudes between 17.0 and 20.0, but the type of variability
is not apparent due to the sparse coverage

\section{ZTF light curve}

   Using the Zwicky Transient Facility (ZTF: \cite{ZTF})
public data\footnote{
   The ZTF data can be obtained from IRSA
$<$https://irsa.ipac.caltech.edu/Missions/ztf.html$>$
using the interface
$<$https://irsa.ipac.caltech.edu/docs/program\_interface/ztf\_api.html$>$
or using a wrapper of the above IRSA API
$<$https://github.com/MickaelRigault/ztfquery$>$.
}, I found that this object was in a standstill
at least between 2021 February and September
(T. Kato on 2022 March 13, vsnet-chat 9014\footnote{
  $<$http://ooruri.kusastro.kyoto-u.ac.jp/mailarchive/vsnet-chat/8457$>$.
}).  Considering that this object is an eclipsing
cataclysmic variable in the period gap, this finding
is surprising.  I here analyze the data in more details.

\begin{figure*}
\begin{center}
\includegraphics[width=16cm]{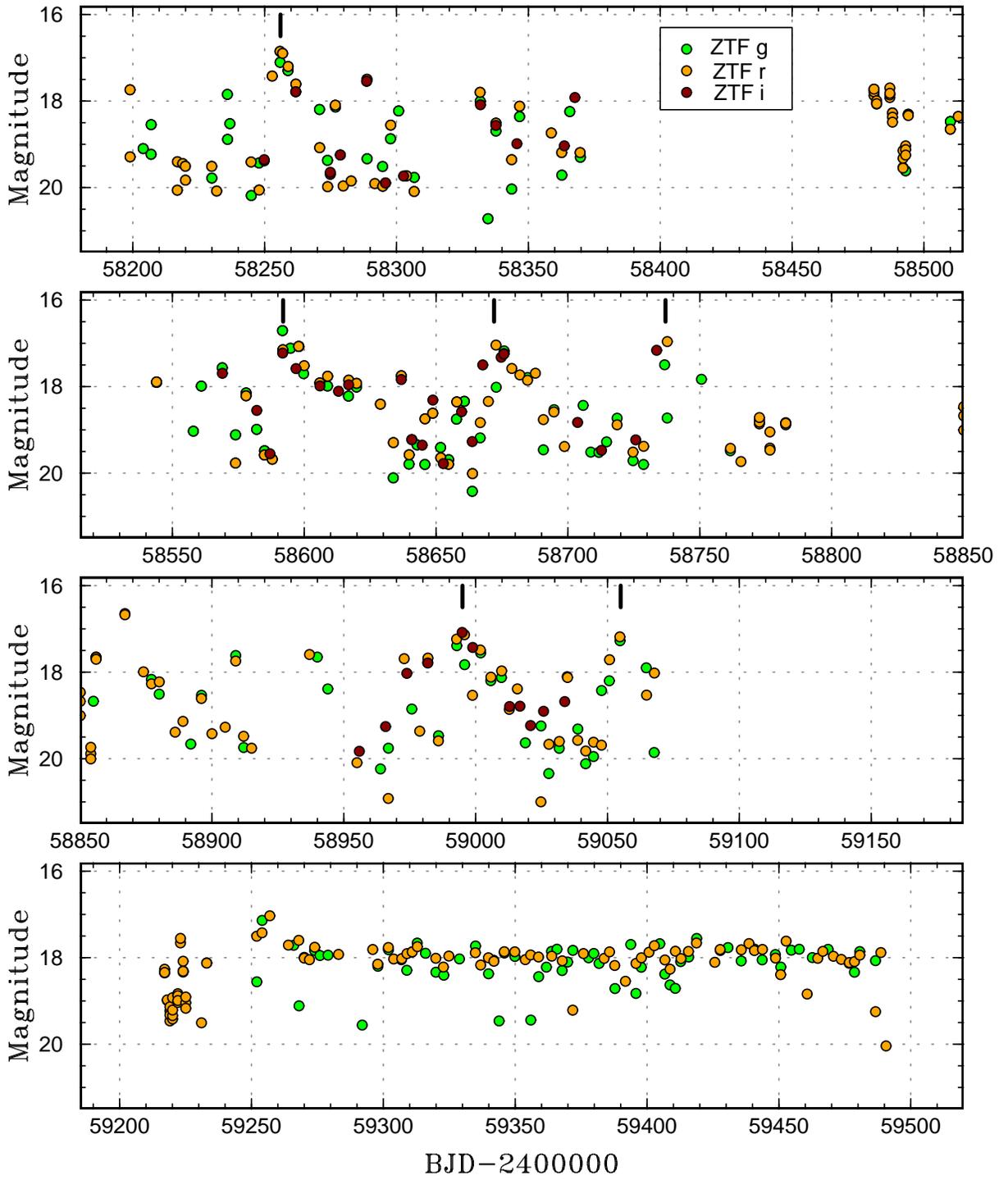}
\caption{
  ZTF light curve of WFI J161953.3$+$031909.
  The object was initially in dwarf nova-type state.
  After BJD 2459260, it entered a well-defined, long standstill.
  The sporadic excursions to fainter magnitudes represent
  eclipses.
  The vertical ticks in the upper three panels represent
  likely superoutbursts.
}
\label{fig:j1619cyglc}
\end{center}
\end{figure*}

   The entire ZTF light curve is shown in
figure \ref{fig:j1619cyglc}.  The object was initially in
dwarf nova-type state (in 2018--2020).
After BJD 2459260 (2021 February), it entered a well-defined,
long standstill.  The object is now confirmed to be
a Z Cam star [for general information of cataclysmic variables
and dwarf novae, see e.g. \citet{war95book}]
in the period gap.  There is only another known
Z Cam star in the period gap (NY Ser, \cite{kat19nyser}),
which will be discussed later.  WFI J161953.3$+$031909
is the first eclipsing Z Cam star in the period gap.
There is (yet) no indication of the IW And-type phenomenon
in this object [see e.g. \citet{sim11zcamcamp1,kat19iwandtype} for
IW And-type stars].

\section{Orbital period}

\begin{figure*}
  \begin{center}
    \includegraphics[width=16cm]{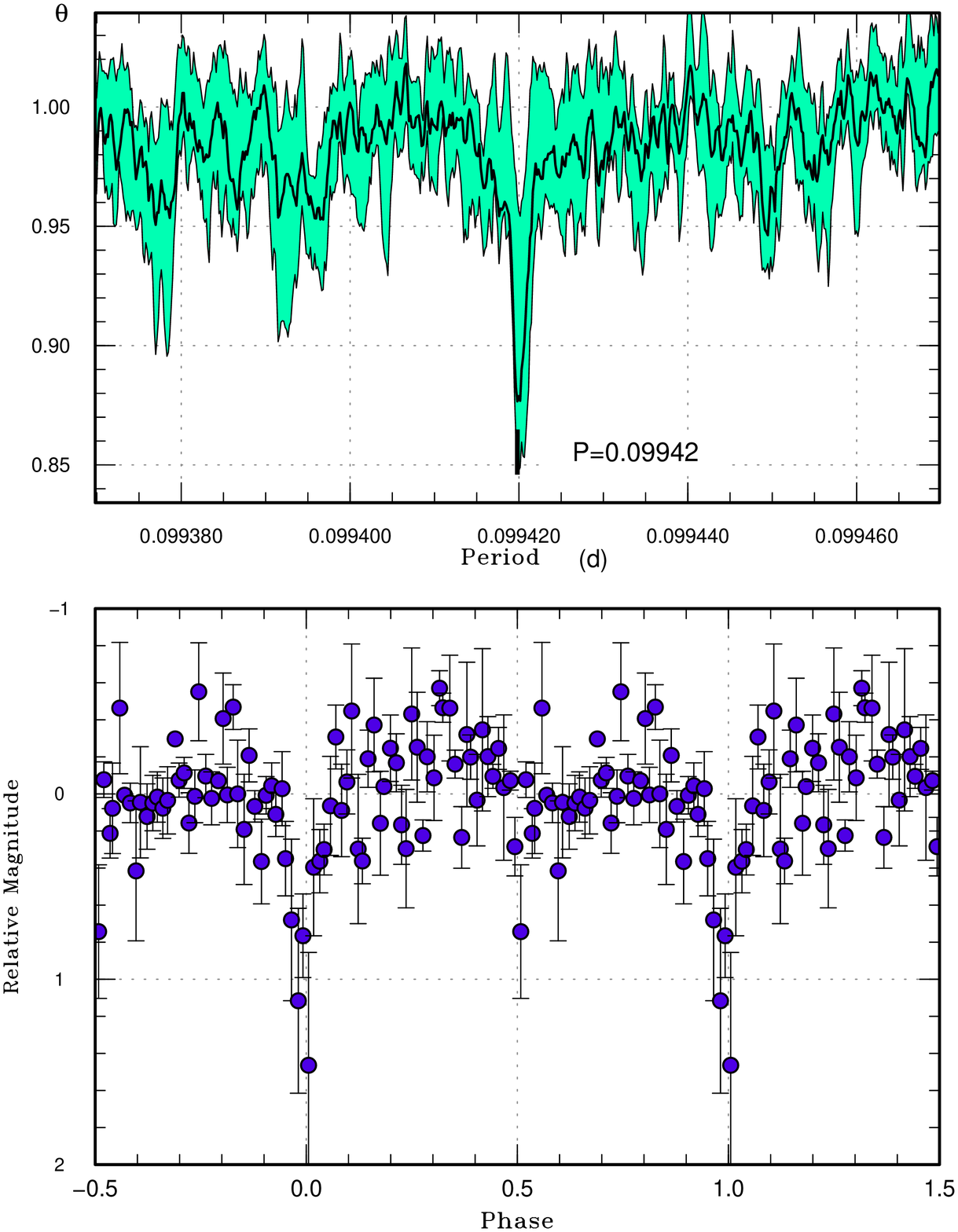}
  \end{center}
  \caption{PDM analysis of WFI J161953.3$+$031909 using the ZTF data.
  (Upper): PDM analysis.
  (Lower): mean profile.  The epoch and period were from
  the MCMC analysis.
  }
  \label{fig:j1619porb}
\end{figure*}

   Due to the presence of both dwarf nova-type variations
and eclipses, and due to the sparse sampling, period analysis of
the ZTF data is rather difficult.  I smoothed $g$, $r$ and $i$
observations separately by locally-weighted polynomial
regression (LOWESS: \cite{LOWESS}) using global smoothing
parameters of $f$=0.05 to reduce the effect of dwarf nova
outbursts.  I then combined them into
a single data set and performed phase dispersion minimization
(PDM: \cite{PDM}) analysis.  The upper panel of
figure \ref{fig:j1619porb} shows the result of
the PDM analysis.  There was no signal other the indicated
one around the period by \citet{rau07DNe}.
The orbital period was refined using
the Markov-Chain Monte Carlo (MCMC)-based
method introduced in \citet{Pdot2}.
The resultant ephemeris is
\begin{equation}
{\rm Min (BJD)} = 2458790.81347(4) + 0.099419808(8) E.
\label{equ:j1619}
\end{equation}
The averaged orbital light curve based on this
ephemeris is shown in the lower panel of
figure \ref{fig:j1619porb}.  Note that this light curve
contains observations in all states (dwarf nova outbursts,
quiescence and standstill).  The result, however,
is very similar to the one in \citet{rau07DNe} without
a pronounced orbital hump.

   Since \citet{rau07DNe} reported no variations outside
the eclipses on seven epochs in 2006 May, it was likely
\citet{rau07DNe} observed the object during a standstill
similar to the one in 2021.  The brightness ($R$=17.4--17.8)
outside the eclipses also supports this idea.  I cannot,
however, perfectly exclude a possibility that they
observed the object during the late phase of a superoutburst
when superhumps became less apparent.
In either case, the object was observed when the accretion
disk was hot.  This explains why there was no pronounced
orbital hump, on the contrary to the explanation by
\citet{rau07DNe} considering a low mass-transfer rate.
By looking back using the ephemeris (\ref{equ:j1619}),
the epoch listed in \citet{rau07DNe} was offset by
$\sim$0.17 phase.  I did not attempt to link the ephemerides
since the orbital period in this study was derived from
observation far from ideal to determine the period and
the actual error may be much larger than the nominal
one shown in equation (\ref{equ:j1619}).

\section{WFI J161953.3$+$031909 as a likely ER UMa star}

\begin{figure*}
\begin{center}
\includegraphics[width=16cm]{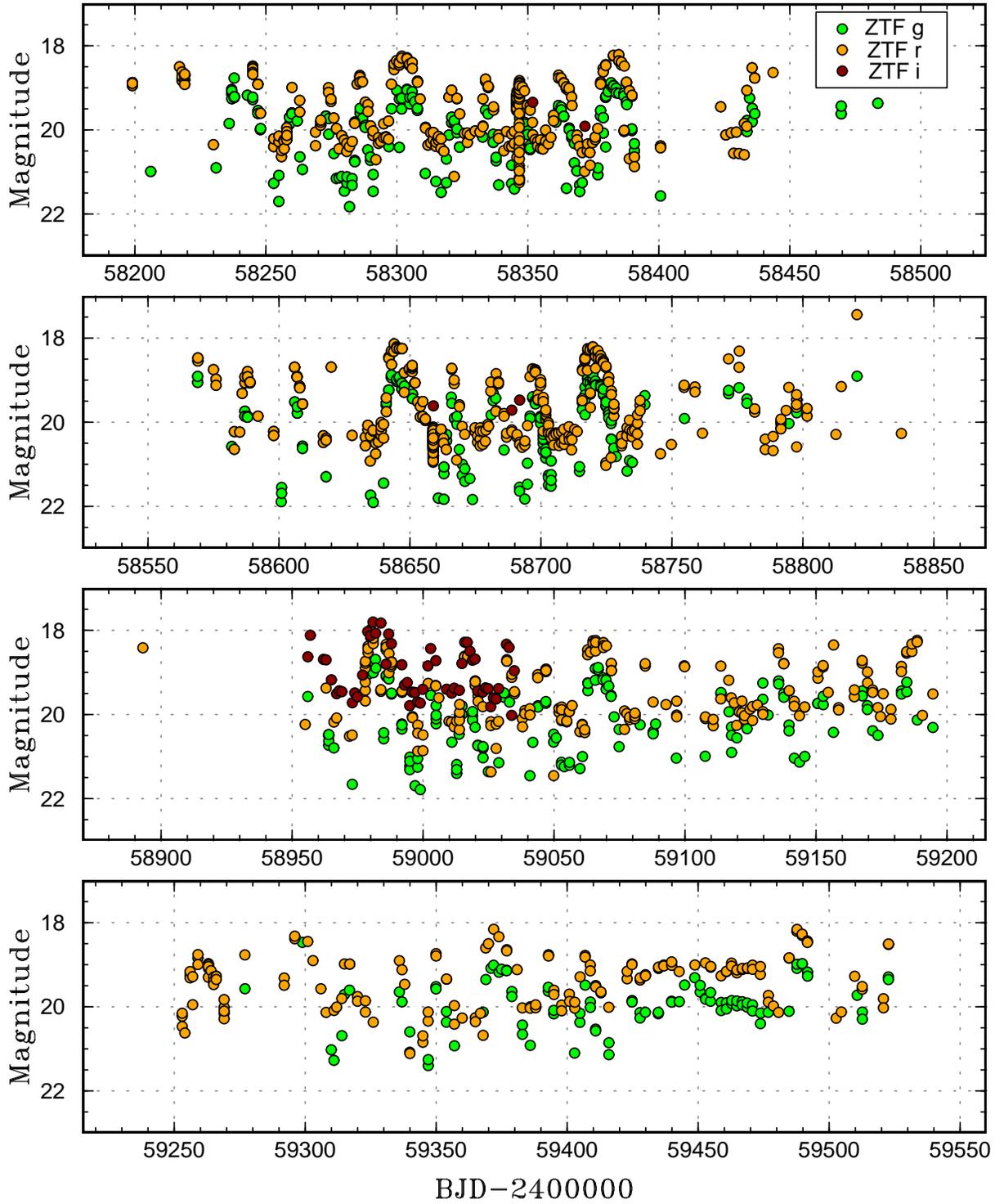}
\caption{
  ZTF light curve of BMAM-V383 = IPHAS J200822.55$+$300341.6.
  The light curve is characterized by long and short outbursts,
  which are usually seen in SS Cyg stars.
  The presence of a standstill terminated by brightening
  (IW And-type phenomenon) is apparent in
  BJD 2459423--2459492.
}
\label{fig:j2008lc}
\end{center}
\end{figure*}

   During the dwarf nova-type state in 2018--2020,
WFI J161953.3$+$031909 showed long outbursts, which are
marked with vertical ticks in figure \ref{fig:j1619cyglc}.
Considering the short orbital period, these long outbursts
must have been superoutbursts of an SU UMa star.
The longest outburst (the first one in the second panel
of figure \ref{fig:j1619cyglc}) lasted at least 28~d,
and possibly 37~d.  The intervals of successive superoutbursts
in the second panel of figure \ref{fig:j1619cyglc}
were 80~d and 65~d.  The interval of the successive
superoutburst in the third panel of figure \ref{fig:j1619cyglc}
was 60~d.  The shortness of the supercycle
(interval between successive superoutbursts) and the long
duration of superoutbursts (comprising 35--46\% of
the supercycle in the extreme case) are perfectly compatible
with the traditional ER~UMa-type classification
[see \cite{kat95eruma}; \cite{rob95eruma}; \cite{pat95v1159ori}; 
for a review, see \cite{kat99erumareview}].
Although BMAM-V383 = IPHAS J200822.55$+$300341.6
\citep{IPHAS2} is listed as an eclipsing ER UMa star
in the AAVSO VSX (at the time of present writing),
the slow fading rates from outbursts
(figure \ref{fig:j2008lc}; see also the light curve of
the SS Cyg star CY Lyr in \cite{Pdot5}) are incompatible with
the ER~UMa-type classification.
This object showed a standstill in
2019 July--September with brightening at the end
(figure \ref{fig:j2008lc}) and it should be re-classified as
an IW And star.  Although its orbital period is not
yet known, the figure\footnote{
  $<$https://www.aavso.org/vsx\_docs/838271/3635/BMAM-V383\%20View\%20on\%20eclipse.PNG$>$.
} presented by Mariusz Bajer
apparently shows a period longer than the period gap,
compatible with an IW And star, but not with
an ER UMa star. 
V4140 Sgr is the only known eclipsing ER UMa-like star
\citep{kat18v4140sgr}.  If superhumps are confirmed
during superoutbursts of WFI J161953.3$+$031909,
this will become the second case of eclipsing
ER UMa stars.

   If superhumps are confirmed (still not yet,
but are very likely), WFI J161953.3$+$031909
is the third object showing both characteristics
of Z Cam and SU UMa stars.  The others are NY Ser
in the period gap \citep{kat19nyser} and BO Cet,
which is also an IW And star, above the period gap
\citep{kat21bocet}.  NY Ser is particularly similar
to WFI J161953.3$+$031909 in that both show frequent
superoutbursts, which is a consequence of a high
mass-transfer rate despite that they are in
the period gap.  At least two standstills in NY Ser
were terminated by superoutbursts \citep{kat19nyser},
which confirmed that the radius of an accretion disk
can increase during standstills and this finding led to
a working hypothesis that the IW And-type phenomenon
is a manifestation of the increase in the disk radius
during standstills (\cite{kat19nyser}; \cite{kim20kic9406652};
M. Shibata et al. in preparation).
In the case of WFI J161953.3$+$031909, we can directly
determine the evolution of the structure of the disk
by using eclipses.  We can expect to learn from
WFI J161953.3$+$031909 what is actually happening
in the ER UMa-type and Z Cam-type disk, and potentially
about the IW And-type disk if the star exhibits
the IW~And-type phenomenon in future.

\section*{Acknowledgements}

This work was supported by JSPS KAKENHI Grant Number 21K03616.
The author is grateful to the ZTF team
for making their data available to the public.
We are grateful to Naoto Kojiguchi for
helping downloading the ZTF data.
This research has made use of the AAVSO Variable Star Index
and NASA's Astrophysics Data System.

Based on observations obtained with the Samuel Oschin 48-inch
Telescope at the Palomar Observatory as part of
the Zwicky Transient Facility project. ZTF is supported by
the National Science Foundation under Grant No. AST-1440341
and a collaboration including Caltech, IPAC, 
the Weizmann Institute for Science, the Oskar Klein Center
at Stockholm University, the University of Maryland,
the University of Washington, Deutsches Elektronen-Synchrotron
and Humboldt University, Los Alamos National Laboratories, 
the TANGO Consortium of Taiwan, the University of 
Wisconsin at Milwaukee, and Lawrence Berkeley National Laboratories.
Operations are conducted by COO, IPAC, and UW.

The ztfquery code was funded by the European Research Council
(ERC) under the European Union's Horizon 2020 research and 
innovation programme (grant agreement n$^{\circ}$759194
-- USNAC, PI: Rigault).

We acknowledge ESA Gaia, DPAC and the Photometric Science
Alerts Team (http://gsaweb.ast.cam.ac.uk/\hspace{0pt}alerts).

\section*{List of objects in this paper}
\xxinput{objlist.inc}

\section*{References}

We provide two forms of the references section (for ADS
and as published) so that the references can be easily
incorporated into ADS.

\renewcommand\refname{\textbf{References (for ADS)}}

\newcommand{\noop}[1]{}\newcommand{\hyphalt}{-}

\xxinput{j1619aph.bbl}

\renewcommand\refname{\textbf{References (as published)}}

\xxinput{j1619.bbl.vsolj}

\end{document}